\documentclass[11pt]{article}
\usepackage[utf8]{inputenc}
\usepackage[T1]{fontenc}
\usepackage{amsmath,amssymb}
\usepackage[letterpaper, margin=1in]{geometry}
\usepackage[backend=biber,style=apa]{biblatex}
\usepackage{graphicx}
\usepackage{booktabs}
\usepackage{longtable}
\usepackage{tabularx}
\usepackage[normalem]{ulem}
\usepackage{parskip}
\usepackage{xcolor}
\usepackage[hidelinks,unicode]{hyperref} 
\usepackage{xurl}
\usepackage[affil-it]{authblk}
\usepackage[symbol]{footmisc}

\title{Toward a Global Regime for Compute Governance: Building the Pause Button}
\author[1]{Ananthi Al Ramiah}
\author[1]{Raymond Koopmanschap}
\author[1]{Josh Thorsteinson}
\author[1]{Sadruddin Khan}
\author[1]{Jim Zhou}
\author[1]{Shafira Noh}
\author[2]{Joep Meindertsma}
\author[3]{Farhan Shafiq\thanks{AI Safety Camp Project Lead}}
\affil[1]{Independent}
\affil[2]{PauseAI}
\affil[3]{Think Safe AI}
\date{\today}

\begin{document}
\maketitle

\begin{abstract}
As AI capabilities rapidly advance, the risk of catastrophic harm from large-scale training runs is growing. Yet the compute infrastructure that enables such development remains largely unregulated. This paper proposes a concrete framework for a global \emph{Compute Pause Button}: a governance system designed to prevent dangerously powerful AI systems from being trained by restricting access to computational resources. We identify three key intervention points—technical, traceability, and regulatory—and organize them within a Governance--Enforcement--Verification (GEV) framework to ensure rules are clear, violations are detectable, and compliance is independently verifiable. Technical mechanisms include tamper-proof FLOP caps, model locking, and offline licensing. Traceability tools track chips, components, and users across the compute supply chain. Regulatory mechanisms establish constraints through export controls, production caps, and licensing schemes. Unlike post-deployment oversight, this approach targets the material foundations of advanced AI development. Drawing from analogues ranging from nuclear non-proliferation to pandemic-era vaccine coordination, we demonstrate how compute can serve as a practical lever for global cooperation. While technical and political challenges remain, we argue that credible mechanisms already exist, and that the time to build this architecture is now, before the window for effective intervention closes.
\end{abstract}

\section{Executive Summary}\label{sec:exec_summary}
As AI capabilities accelerate, the risks associated with large-scale
training runs are growing rapidly. The most powerful AI systems ---
those capable of triggering profound societal disruption or catastrophic
harm --- require vast amounts of compute. These risks range from mass
disinformation, cyberattacks, and biological/chemical weapon
acceleration to economic destabilization, gradual loss of human agency,
and, in the extreme, existential risk through permanent loss of human
control. Yet current governance efforts overwhelmingly focus on AI use
and deployment, leaving the infrastructure that enables dangerous
training runs largely unregulated. According to a
recent estimate \parencite{kokotajloAI20272025}, at current rates
of progress, frontier labs could cross critical danger thresholds as
early as 2027--2028. Without credible mechanisms to pause or prevent
threshold-exceeding compute, we face the very real possibility of losing
control over AI development.

This report proposes a concrete, enforceable framework for building a
global Compute Pause Button: a governance system that prevents dangerous
AI training runs by restricting access to compute. It targets three key
intervention points:

1. Technical --- controlling compute through hardware modifications
(e.g., modifying hardware to enforce FLOP ceilings, limiting
interconnectivity, restricting model deployment);

2. Traceability --- creating end-to-end traceability infrastructure
(e.g.,tracking chips, components and users to ensure compute flows
remain visible and auditable);

3. Regulatory --- establishing standards and legal architecture (e.g.,
through export controls, production quotas, licensing schemes).

For each intervention, we apply an integrated Governance, Enforcement,
and Verification (GEV) framework that makes rules clear, violations
detectable, and compliance independently verifiable. Mechanisms such as
tamper-proof FLOP caps, offline licensing, model locking, fixed-set
configurations, chip registries, and KYC protocols illustrate how this
framework can be operationalized in practice. By meeting GEV criteria,
these mechanisms create a layered and resilient system of oversight
that offers a credible blueprint for halting the development of
dangerously powerful AI systems before deployment.

In shifting governance towards policies and mechanisms that prevent
these powerful models from being trained and deployed in the first
place, our proposal would make dangerous development not only illegal but logistically and technically infeasible. This approach draws
inspiration from real-world analogues such as the Nuclear
Non-proliferation Regime (NPT/IAEA), the Chemical Weapons Convention
(CWC/OPCW), and the Wassenaar Arrangement on export controls. Together,
they show how dangerous technologies can be governed through a
combination of legal instruments, verification systems, and
institutional cooperation.

This report does not assume that political will, institutional
readiness, or technical maturity are already in place. Rather, it aims
to lay out what a pause requires, and by showing that credible
intervention points do exist. We acknowledge that significant work
remains, from developing technical safeguards to building international
consensus, but we believe that the proposed architecture is sound, and
the window for action is narrowing.

We call on policymakers, chipmakers, AI labs, and multilateral bodies to
begin coordinated planning around compute governance now. The Pause
Button is not a fantastical notion. It is a necessary safeguard, and one
that must be engineered before dangerous systems become ungovernable.
With credible interventions and international cooperation, we can act
before the window for effective intervention closes, possibly within
just a few years. The time to act is now.

\hypertarget{introduction}{%
\section{Introduction}\label{introduction}}

The accelerating capabilities of AI systems --- and the potential for
catastrophic and irreversible harms \parencite{centerforaisafetyStatementAIRisk2023} --- have made the need for
international agreements to regulate or even
pause frontier AI development increasingly urgent \parencite{russellHumanCompatibleArtificial2020, futureoflifeinstitutePauseGiantAI2023}. Advanced models could
soon enable mass disinformation, critical infrastructure attacks,
biological weapon acceleration, or even the loss of meaningful human
oversight. According to a recent estimate \parencite{kokotajloAI20272025}, at current rates
of progress, frontier labs could cross critical danger thresholds as
early as 2027--2028. Yet, while calls for
international agreements have grown louder, the concrete
mechanisms \parencite{aguirreChapter8How2025}
and frameworks \parencite{scholefieldInternationalAgreementsAI2025} required
to enact such a pause remain underdeveloped. Recent
work has explored the
diplomatic and institutional
groundwork and infrastructure required to create an
international agreement to stop dangerous AI development \parencite{wasilVerificationMethodsInternational2024, juijnAdvancedAITechnical2024, maasInternationalAIInstitutions2023a}. This white paper is
complementary to --- and builds up from --- such recent work. Our focus
is on developing a technical and regulatory framework and infrastructure
that would prevent AI training runs beyond a critical compute threshold;
in effect, constructing a metaphorical, yet implementable,
\textbf{``Compute Pause Button.''}

Compute refers to the processing power required to train or run AI models \parencite{buchananAITriadWhat2020},
typically measured in FLOPs (floating-point operations). In AI
development, especially for large-scale models like GPT-4, total compute
(FLOPs used during training) is a key determinant of model capabilities.
In the last fourteen years, the computing power used to train top AI
models has grown by a staggering
1.5 billion times \parencite{EpochNotableModels2024}; an exponential
rise that underpins many of today's most notable AI breakthroughs \parencite{pilzIncreasedComputeEfficiency2023}.

This white paper reviews existing proposals for limiting large-scale AI
training runs and organizes them within a unified governance framework.
Our vision for the Compute Pause Button is to establish a globally
enforceable policy regime that prevents the development of dangerously
advanced AI systems. This would be achieved either by restricting access
to the compute infrastructure required for such training (e.g., through
supply chain controls, licensing, and export controls) or by regulating
the deployment and configuration of the hardware itself, ensuring, for
example, that authorized actors or automated mechanisms can operate
datacenter-scale clusters when they exceed predefined compute
thresholds.

The rest of this document is organized as follows:

\begin{itemize}
\item
  \hyperref[pause-thresholds]{Section 3} discusses
  different threshold metrics that could trigger a pause in AI
  development;
\item
  \hyperref[intervention-points]{Section 4} outlines
  our proposed intervention points;
\item
  \hyperref[governance-framework-to-enact-a-pause]{Section 5} presents a governance framework for putting them into practice;
\item
  \hyperref[translating-intervention-points-into-action-mechanisms-and-their-gev-pathways]{Section 6} details the specific mechanisms for each intervention point, organized within the GEV framework;
\item
  \hyperref[relevant-historical-policy-analogues-precedents-for-global-coordination-and-the-need-for-institutional-imagination]{Section 7} distills key lessons from relevant policy analogues to inform our approach;
\item
  \hyperref[conclusion-engineering-the-pause-button-together]{Section 8} addresses the limitations of this work and concludes the report.
\end{itemize}

\hypertarget{pause-thresholds}{%
\section{Pause Thresholds}\label{pause-thresholds}}

Assessing the risk of catastrophic AI failure is inherently difficult.
Unlike aviation, where risk is precisely quantifiable through historical
data and known failure modes, advanced AI systems are novel, opaque, and
rapidly evolving across many domains. Their potential failure modes are
not only poorly understood but may be fundamentally unpredictable. One
potential approach is model evaluations, which aim to directly assess
the dangerous capabilities and propensities of AI systems. Yet dangerous
capability evaluations fail to address unknown failure modes.

This white paper advocates for compute thresholds as a pragmatic, albeit
indirect, proxy for AI risk. Focusing on compute targets a quantifiable
input that is essential for training frontier models. This approach
offers strong objectivity, builds on existing policy discussions (such
as FLOP-based thresholds in the EU AI Act and the former US Executive
Order on AI), and enables early intervention --- before potentially
uncontrollable capabilities emerge.

Floating-point operations (FLOP) and floating-point operations per
second (FLOPS) have emerged as key metrics for compute governance. FLOP
quantifies the number of calculations invested in a training run,
serving as a proxy for its potential capabilities. FLOPS, by contrast,
measures the processing speed of hardware or a compute cluster,
indicating its capacity for large-scale training. Regulatory proposals
often cite specific figures --- e.g., 10\textsuperscript{20} FLOPS for
cluster capacity or 10\textsuperscript{25} FLOP as an initial ceiling,
which, if crossed, trigger higher scrutiny.

While FLOP-based metrics are strong as a primary trigger, a
comprehensive understanding of a system\textquotesingle s practical
training capacity also requires consideration of other factors. Memory
operations per second (MOPS) and the resulting arithmetic intensity
(defined as FLOPS divided by MOPS) help determine whether a system is
limited by compute or memory, highlighting the
memory wall bottleneck that can limit the effective utilization of theoretical FLOPS \parencite{10.1145/216585.216588, gholamiAIMemoryWall2024}.
Similarly, interconnect bandwidth, the speed at which data can be shared
between processing units, is crucial for large distributed training
clusters. While these metrics are useful for detailed system analysis
and could inform secondary checks or investigations, their complexity,
and variability make them less suitable than FLOP or FLOPS as a clear
and consistent trigger for a pause.

A key challenge in using a compute threshold is that, over time,
continued advancements in algorithmic
efficiency and other forms of technological progress will decrease the
compute required to develop highly capable AI systems \parencite{ho2024algorithmic}. Innovations in
model architecture and training techniques would soon render any fixed
threshold outdated. To account for this, a low-tech method is to
regularly convene an expert panel to revise these limits. Given the
speed of innovation in AI, regulators should be empowered to revise
thresholds quickly. The EU AI Act does
this by allowing the Commission to issue delegated acts updating the
FLOP thresholds and adding new benchmarks without full Parliament
approval \parencite[58]{pistillorole}.

However, this doesn't address how an expert panel would choose a new
threshold. For this, one available tool is capability evaluations.
Above, we argued that evaluations are inappropriate as the primary
threshold itself because they cannot account for every potential risk
posed by an advanced AI system. However, it may be feasible to sidestep
this issue using a suite of broad capability evaluations rather than
aiming to pinpoint the emergence of dangerous capabilities. The
threshold can be adjusted when a model reaches a given level of
performance using fewer computational resources than the previous state
of the art. While evaluations are not sufficient as a direct primary
trigger for a pause due to unknown risks, they are valuable for
calibrating the compute threshold itself over time against known
capability benchmarks. Further research is needed on the details of this
plan. For example, which capability evaluations would be appropriate?
And until the science of evaluations progresses \parencite{apolloresearchWeNeedScience2024}, a substantial safety margin to account for
their imprecision would be necessary.

\hypertarget{intervention-points}{%
\section{Intervention Points}\label{intervention-points}}

The Compute Pause Button is our proposed strategy to temporarily halt
the training of AI models beyond a certain computational threshold, in
an effort to prevent the development of uncontrollably powerful AI. It
relies on technical, traceability, and regulatory tools (which we call
intervention points) to turn the metaphor of a Compute Pause Button into
a concrete, enforceable, and verifiable system of control and
management. These intervention points are the places within which policy
levers or mechanisms can be applied to restrict and monitor the use of
high-powered compute. Together, these three intervention points form the
structural foundation on which any pause would rest. These three
intervention points are like legs of a tripod; if you remove one, the
system is liable to collapse. A credible Compute Pause Button requires
all three to work together: technical containment, traceability
infrastructure, and regulatory force (examples of the intervention
points are provided in Table 1).

\begin{table}[h]
\centering
\caption{Intervention Points and Their Mechanisms for Implementation}
\label{tab:intervention_points_fit}
\begin{tabularx}{\linewidth}{@{} 
  >{\raggedright\arraybackslash}p{0.5\linewidth} 
  >{\raggedright\arraybackslash}X 
  @{}}
\toprule
\textbf{Intervention Points} & \textbf{Mechanisms} \\
\midrule
Technical: Controlling Compute Through Hardware Modifications (e.g., modifying hardware to enforce FLOP ceilings, limiting interconnectivity, restricting model deployment) & 
\hyperref[tamper-proof-flop-caps-enforcing-hard-limits-on-compute-usage-at-the-hardware-level]{FLOP Caps}, 
\hyperref[offline-licensing-enforcing-compute-limits-through-time-limited-access-controls]{Offline Licensing}, 
\hyperref[fixed-set-mechanisms-preventing-large-compute-clusters-by-capping-ai-chip-networking-capabilities]{Fixed Set Mechanisms}, 
\hyperref[model-locking-preventing-the-unauthorized-deployment-of-model-weights-outside-of-a-secured-environment]{Model Locking} \\
\midrule
Traceability: Creating End-to-End Traceability Infrastructure (e.g., tracking chips, components and users to ensure compute flows remain visible and auditable) & 
\hyperref[monitoring-key-materials-and-chip-components-preventing-illicit-compute-production]{Monitoring Key Materials}, 
\hyperref[global-compute-supply-registry-requiring-all-chip-production-facilities-to-register-with-a-central-verification-body-to-log-volumes]{Global Compute Supply Registry}, 
\hyperref[know-your-customer-kyc-schemes-vetting-access-to-large-scale-compute-resources]{KYC}, 
\hyperref[chain-of-custody-framework-for-high-performance-chips-tracking-the-ownership-and-movement-of-finished-chips-to-prevent-diversion-or-misuse]{Chain of Custody} \\
\midrule
Regulatory: Establishing Standards and Legal Architecture (e.g., through export controls, production quotas, licensing schemes) & 
\hyperref[export-controls-regulating-the-specific-actors-and-countries-that-can-access-which-chips-and-how-many]{Export Controls}, 
\hyperref[production-controls-limiting-the-volume-and-type-of-high-performance-chips-via-quotas-and-licensing]{Production Controls} \\
\bottomrule
\end{tabularx}
\end{table}

The eleven mechanisms outlined in Table 1 are drawn primarily from a set
of proposals developed by researchers at MIRI,
focused on how to verify and enforce international agreements on
advanced AI \parencite{scher2024mechanisms}. Our selection reflects a design logic in which each
mechanism targets a chokepoint in the global compute supply chain. These
chokepoints (at the point of chip fabrication, hardware distribution,
and/or user access) represent tangible leverage points where policy and
technical controls can be meaningfully applied. Rather than relying
solely on end-use regulation or oversight at the point of AI deployment,
these mechanisms aim to intervene earlier in the development pipeline,
before powerful AI systems are trained. By focusing on
infrastructure-level and supply-side controls, these mechanisms provide
a promising foundation for building an enforceable pause. While not all
mechanisms are deployment-ready (particularly those involving hardware
modifications) they collectively sketch a coherent architecture that,
with further research and investment, could become logistically
feasible.

\hypertarget{technical-intervention-point-controlling-compute-through-hardware}{%
\subsection{\texorpdfstring{Technical Intervention Point:
Controlling Compute Through Hardware
}{Technical Intervention Point: Controlling Compute Through Hardware }}\label{technical-intervention-point-controlling-compute-through-hardware}}

This intervention point focuses on embedding restrictions and
constraints directly into compute hardware. By constraining how chips
are used and interconnected, policymakers can place hard physical and
software-based limits on training runs, which would make unauthorized
training runs technically impossible or prohibitively difficult, thereby
closing off the path to dangerous scale-ups. For example, the hardware
could be configured with the ability to remotely or automatically
shutdown large-scale clusters that exceed a given threshold. Some
mechanisms or policy levers that form the basis of a technical
intervention point would be: FLOP caps (compute ceilings), offline
licensing systems, model locking, and fixed chip cluster configurations.

\hypertarget{traceability-intervention-point-creating-end-to-end-traceability-infrastructure}{%
\subsection{Traceability Intervention Point: Creating End-to-End
Traceability
Infrastructure}\label{traceability-intervention-point-creating-end-to-end-traceability-infrastructure}}

Traceability ensures that compute can be tracked from the production and
distribution of critical components, such as advanced chips and
specialized hardware, through to the organizations capable of
concentrating and applying that compute toward frontier model
development. This intervention point builds the visibility needed for
oversight, enforcement, and accountability. By making compute traceable,
this infrastructure reveals usage that exceeds pre-agreed limits,
prevents black-market flows, and supports both compliance and
deterrence. Some mechanisms or policy levers that form the basis of a
traceability intervention point would be: tracking of critical equipment
and components, creating a global compute registry, chain of custody for
chips, and Know Your Customer (KYC) protocols applied to compute buyers
and post-sale location verification.

\hypertarget{regulatory-intervention-point-establishing-standards-and-legal-architecture}{%
\subsection{\texorpdfstring{Regulatory Intervention Point:
Establishing Standards and Legal Architecture
}{Regulatory Intervention Point: Establishing Standards and Legal Architecture }}\label{regulatory-intervention-point-establishing-standards-and-legal-architecture}}

This intervention point focuses on building the rules, norms, and global
agreements that enable the levers within the technical and traceability
intervention points to function. It defines how much compute is in the
market, who can produce, sell, or access advanced compute, and under
what terms. It authorizes restrictions and provides legitimacy and
enforcement power for a global pause. Some mechanisms or policy levers
that form the basis of a regulatory intervention point would be: export
controls on chips and tools and semiconductor production controls and
quotas.

To ensure these interventions work together effectively to halt
threshold-exceeding training runs, we introduce the Governance,
Enforcement, and Verification (GEV) Framework --- a comprehensive
structure designed to make the Compute Pause Button not just
theoretically sound, but practically enforceable and globally credible.

\hypertarget{governance-framework-to-enact-a-pause}{%
\section{Governance Framework to Enact a
Pause}\label{governance-framework-to-enact-a-pause}}

While the intervention points highlight \emph{\textbf{where}}
intervention would be most effective, the GEV framework outlines the
\emph{\textbf{how}}; it is a functional operating system that enables
governance to be implemented, enforced, and verified over time. GEV
distinguishes between governance (setting the rules), enforcement
(ensuring compliance), and verification (checking whether compliance
occurred).

To effectively govern powerful AI development, compute restriction
policies must unfold both sequentially and cyclically. This framework
draws from established models in
\href{https://www.fatf-gafi.org/en/the-fatf/what-we-do.html}{money
laundering and terrorist financing},
\href{https://www.doit.com/cloud-governance-frameworks-a-comprehensive-guide/}{cloud
governance},
\href{https://www.gao.gov/products/gao-23-106020}{taxation
policy}, and
\href{https://disarmament.unoda.org/wmd/nuclear/npt/\#:~:text=The\%20NPT\%20is\%20a\%20landmark,and\%20general\%20and\%20complete\%20disarmament.}{arms}
\href{https://www.iaea.org/}{control}, which rely on a
structured sequence of rule-setting, compliance mechanisms, and
oversight. To aid understanding, we now explain our GEV framework with
some compute-related examples and also in comparison to the governance,
enforcement, and verification stages involved in common taxation
policies.

\begin{figure}[htbp]
    \centering
    \includegraphics[width=\textwidth]{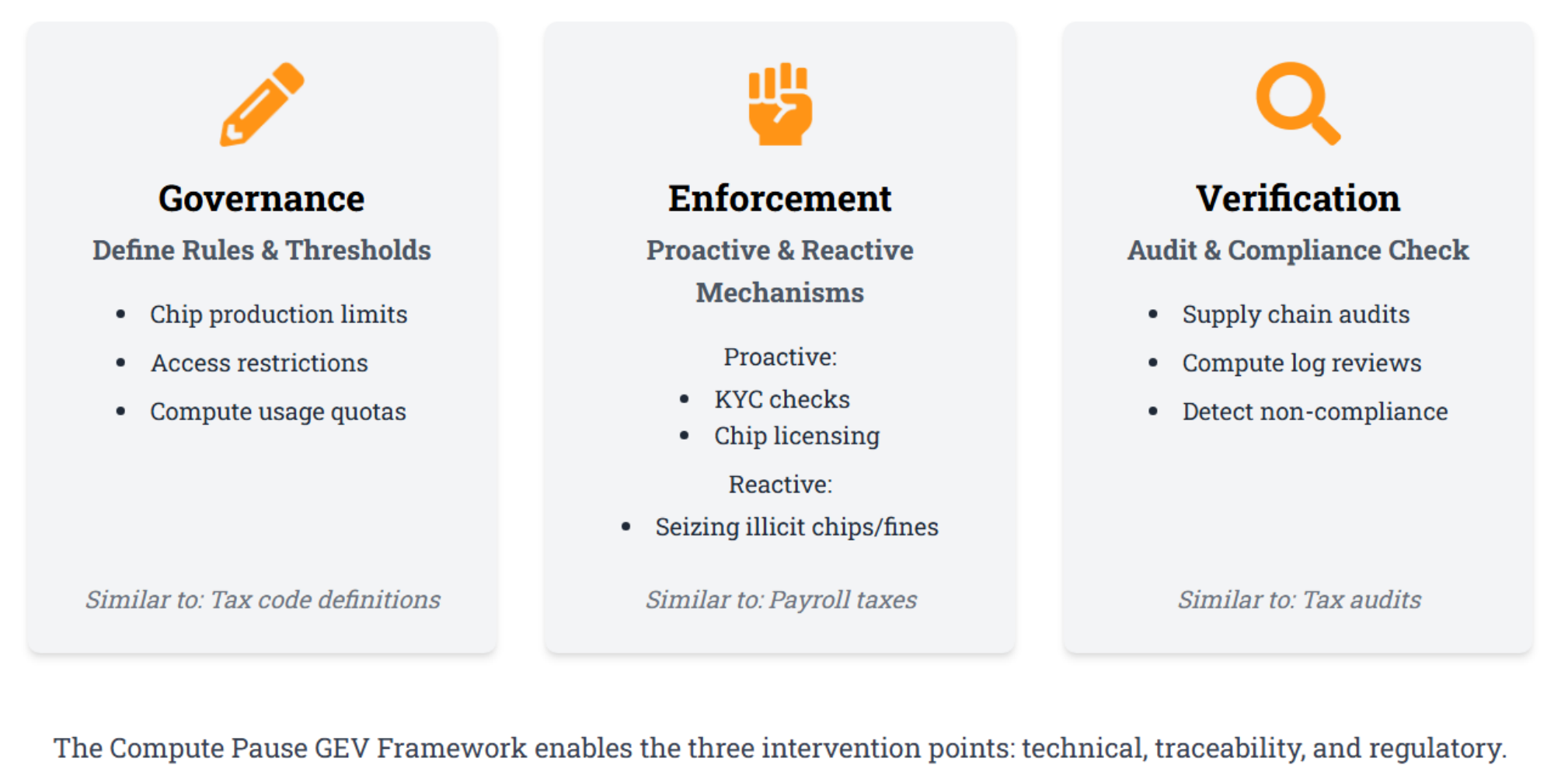}
    \caption{Framework and Example Mechanisms for Compute Pause Button}
    \label{fig:gev_framework}
\end{figure}

First, \textbf{governance} happens before deployment --- e.g., when
rules are written to define who can access and use AI compute and how
many chips can be produced and/or sold. This is similar to tax codes
which define what income is taxable and at what rate. Second,
\textbf{enforcement} happens during AI compute sales and usage, applying
mechanisms such as KYC, licensing, or chip export restrictions that
prevent violations upfront, akin to payroll taxes being withheld before
employees receive wages. But tax enforcement can also involve collecting
back taxes after underpayment is discovered. In other words, enforcement
has proactive (withholding taxes) and reactive (collecting back taxes)
aspects. In terms of compute restrictions, enforcement is proactive when
FLOP caps or chip licensing block unauthorized training runs upfront,
and reactive when regulators seize illicit chips or fine data centers
after detecting violations. Finally, \textbf{verification} is the
after-the-fact process and safeguard, where audits, inspections, and
supply chain reviews ensure the rules are followed, just as tax
agencies audit past financial statements to catch tax evasion and fraud.

Without this multi-component temporal sequencing, any pause on AI
development risks falling apart:

\begin{itemize}
\item
  Governance without enforcement is ignored;
\item
  Enforcement without verification invites covert circumvention.
  Further, verification allows for reactive enforcement as it uncovers
  deviations that may have been missed during proactive enforcement;
\item
  Without clear governance, verification has no benchmark: no way to
  know what counts as noncompliance;
\item
  Verification and enforcement also inform governance in the medium- to
  long-term, as regulations may need to be altered in response to how
  markets are behaving.
\end{itemize}

Thus, each component of GEV reinforces the others, providing both
deterrence and accountability, with the GEV framework unfolding both
sequentially and in a cyclical, reinforcing manner.

\begin{figure}[htbp]
    \centering
    \includegraphics[width=\textwidth]{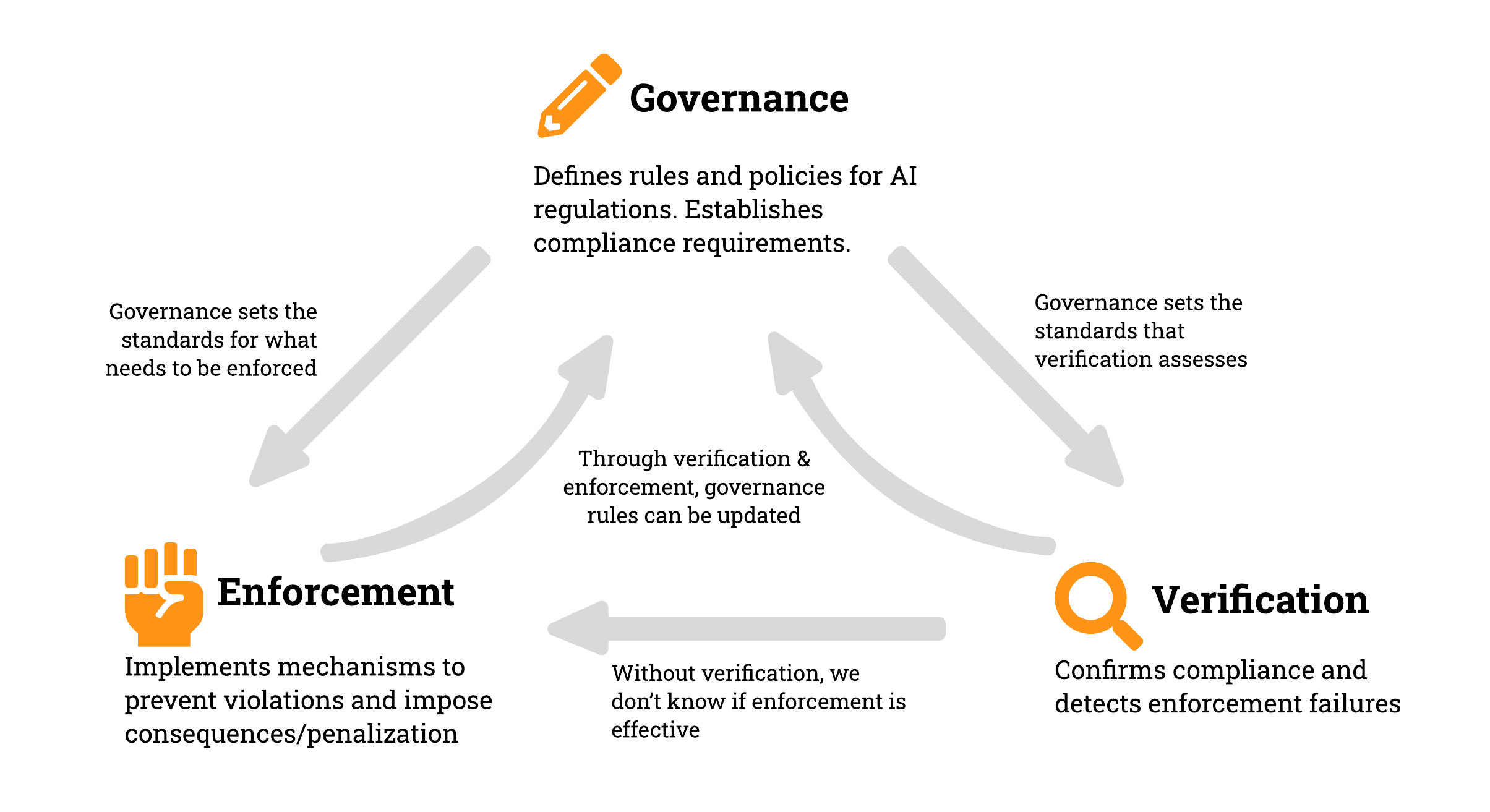}
    \caption{Cyclical nature of GEV}
    \label{fig:gev_distribution}
\end{figure}

The GEV framework makes the Compute Pause Button more than a metaphor:
it becomes a functional, implementable system. Each intervention point
must include mechanisms that fulfill GEV criteria, ensuring rules are
defined, violations are prevented or punished, and compliance can be
verified. Only then can policymakers press the Compute Pause Button and
be confident it will actually work.

Each of the intervention points serves as a critical technical,
traceable, and regulatory avenue through which governance, enforcement,
and verification must be operationalized. The GEV framework activates
the intervention points. For example, governance defines the legal and
institutional authority behind chip licensing or export controls.
Enforcement gives teeth to FLOP caps, KYC protocols, or production
quotas. Finally, verification ensures that traceability infrastructure
and auditing mechanisms actually function as intended. In this way, GEV
threads through and powers each intervention point, transforming
individual mechanisms into a cohesive system capable of reliably pausing
frontier AI development.

\hypertarget{translating-intervention-points-into-action-mechanisms-and-their-gev-pathways}{%
\section{Translating Intervention Points into Action: Mechanisms
and Their GEV
Pathways}\label{translating-intervention-points-into-action-mechanisms-and-their-gev-pathways}}

\hypertarget{technical-intervention-point}{%
\subsection{Technical Intervention
Point}\label{technical-intervention-point}}

The technical intervention point focuses on embedding constraints into
the hardware and firmware that power advanced AI development. Unlike
traceability and regulatory tools, which rely on external monitoring and
legal mandates, technical mechanisms aim to make certain behaviors (such
as training models above a specific compute threshold) physically or
cryptographically infeasible. These mechanisms can operate independently
of a trusted network connection or institutional compliance, offering a
unique pathway to robust, tamper-resistant enforcement.

We will discuss five mechanisms in this section. First, we introduce a
novel hardware solution known as
flexible hardware-enabled guarantees (flexHEGs) \parencite[]{petrie2024interim} which can enforce a
range of limits. We give an example of how it can be used to enforce
hard limits on compute usage. While potentially promising, this option
still needs considerable research and development.

Secondly, we introduce two other mechanisms to enforce limits on
computational resources to detect and prevent unauthorized or
threshold-exceeding training runs. These two mechanisms could be
implemented in multiple ways. We will focus here on the implementation
pathways that do not rely on flexHEGs, as they may be deployable sooner.
While we prioritize near-term implementations, it is important to note
that both mechanisms could also be integrated with flexHEGs in the
future, making them more tamper-resistant.

Finally, we discuss mechanisms designed to prevent trained AI models
from being accessed, transferred, or deployed outside of secure,
authorized environments. All these mechanisms vary in maturity,
tamper-resistance, and ease of implementation. An overview is provided in Table 2.

\begin{table}[p]
\centering
\caption{Technical Mechanisms Overview and Distinctions}
\label{tab:technical_mechanisms}
\vspace{0.5em}
\begin{tabularx}{\linewidth}{@{} 
  >{\raggedright\arraybackslash}X 
  >{\raggedright\arraybackslash}X 
  >{\raggedright\arraybackslash}X 
  >{\raggedright\arraybackslash}X 
  >{\raggedright\arraybackslash}X 
  >{\raggedright\arraybackslash}X 
  @{}}
\toprule
\textbf{Mechanism} & \textbf{Primary purpose} & \textbf{Tamper-resistance} & \textbf{Ease of implementation} & \textbf{Research needed} & \textbf{Downsides} \\
\midrule
\textbf{FlexHEGs} & 
To enable secure, hardware-enforced governance mechanisms & 
High & 
Difficult; requires chip redesign or retrofitting on existing chips & 
Several years \parencite[p. 6]{petrie2024interim} & 
Long research and deployment timeline; not yet deployable \\
\midrule
\textbf{Tamper-proof FLOP Caps} & 
To enforce compute usage ceilings by embedding FLOP limits in hardware & 
High & 
Same as flexHEGs & 
Same as flexHEGs & 
Dependent on flexHEG development; circumvention via task splitting remains a risk \\
\midrule
\textbf{Offline Licensing} & 
To limit compute use through externally validated FLOP quotas & 
Low, when implemented using firmware & 
Easy; can be rolled out via firmware updates on existing chips & 
Within a year \parencite[p. 1]{petrieNearTermEnforcementAI2024a} & 
Vulnerable to circumvention or license forgery if not secured \\
\midrule
\textbf{Fixed Set Mechanisms} & 
To prevent large-scale training by capping chip network abilities (interconnectivity between chips) & 
Medium & 
Medium; implementable through equipment modifications and network programming, though chip-level integration improves reliability & 
$<$1 year \parencite[p. 35]{scher2024mechanisms} & 
Can be partially circumvented by distributed training \\
\midrule
\textbf{Model Locking} & 
To prevent unauthorized deployment of model weights after training & 
High, if implemented with secure hardware & 
Difficult; similar to flexHEGs we need to adapt the chip & 
1--5 years \parencite[p. 51]{scher2024mechanisms} & 
Does not prevent misuse by the original developer; only restricts external misuse \\
\bottomrule
\end{tabularx}
\end{table}

\subsubsection{FlexHEGs: Tamper-proof hardware modules to verify and enforce safety guarantees}
\label{flexhegs---tamper-proof-hardware-modules-to-verify-and-enforce-safety-guarantees}

FlexHEGs are conceptual hardware modules, either integrated on-chip or
attached as a device, that could enable privacy-preserving, trustworthy,
and multilateral enforcement of AI governance protocols. They are
envisioned to support FLOP accounting and enforcement, dataset size
restrictions, model architecture verification, licensing-based usage,
and so forth. If realized, flexHEGs could serve as a foundation for
multiple technical controls across the AI lifecycle, from training to
deployment.

In the flexHEG setup, each powerful AI chip is sealed inside a secure and
tamper-proof box together with a small ``supervisor'' processor that
watches every instruction and data transfer. This overseer processor can
also check and prove, in a privacy-preserving way, exactly what the chip
is doing, thereby ensuring agreed-upon limits are never exceeded and
safety rules can be enforced. They offer great flexibility and
multilateral governance potential.

However, flexHEGs are not yet available, and significant research and
development would be required to make them viable at scale, and
timelines for deployment are uncertain. As such, the offline licensing
and fixed set mechanisms that we discuss now should be seen not only as
candidates for integration with flexHEG, but also as modular
interventions worth pursuing in parallel --- both to reduce near-term
risk and to lay the groundwork for more robust, enforceable governance
in the future.

\subsubsection{Tamper-Proof FLOP Caps: Enforcing Hard Limits on Compute Usage at the Hardware Level}
\label{tamper-proof-flop-caps-enforcing-hard-limits-on-compute-usage-at-the-hardware-level}

Tamper-proof FLOP caps are one of the most direct technical approaches
to enforcing compute limits. The core idea is simple: monitor and
restrict how much computational power (measured in FLOPs) an AI chip can
use, and automatically shut down or block operations once a predefined
threshold is reached.

This mechanism can be implemented in different ways. One approach is to
integrate FLOP monitoring directly into chips at the point of
manufacture. Another option is to retrofit existing chips with secure
hardware modules that track compute use and intervene if limits are
breached. Both approaches can include tamper-resistance features, such
as physically hardened enclosures or self-deactivation if tampering is
detected.

As introduced in the previous paragraph, a tamper-resistant way to implement this mechanism is using flexHEGs. Petrie et al. propose a system where ``each secure processor can keep track of how much computation has gone into producing various intermediate results, such as activations, gradients, and shards of weights, and pass that information to the next processor in a `receipt' accompanying the intermediate result itself. These receipts can then be traced back all the way to the randomly initialized starting point and summed to obtain the total number of floating-point operations that contributed to a final set of model weights'' (\citeyear{petrie2024interim}).

For FLOP caps to work at scale, a \textbf{governance} body (either
international or composed of trusted multilateral actors) would need to
define allowable FLOP thresholds and certify hardware designs that
enforce them. This body would also oversee the authorization of firmware
updates, ensuring that updates don't weaken enforcement or introduce
vulnerabilities. By maintaining an open-source design for both the
secure processor and the tamper-resistant enclosure, all parties can be
assured of the system's integrity and that the FLOP capping measures are
applied uniformly across relevant hardware.

\textbf{Enforcement} is built into the hardware itself. If the system
detects that a training run is about to exceed its authorized compute
limit, it can automatically halt execution. In some configurations, it
may also trigger a shutdown or self-disablement if tampering is
suspected. These interlocks make it costly and difficult for bad actors
to override or bypass FLOP limits. Effective enforcement requires that
the secure processor and its tamper-resistant enclosure be mutually
trusted by all relevant stakeholders to ensure that no vulnerabilities
have been introduced into the system.

\textbf{Verification} would rely on secure logging or recording.
Hardware components --- from simple counters to sophisticated setups
like FlexHEG --- can produce secure digital ``receipts'' that record and
summarize exactly what calculations they've carried out. These receipts
can be audited periodically or in real-time to confirm that no system
has exceeded its limit. However, an important limitation is that FLOP
caps applied at the level of individual chips or devices could, in
theory, be circumvented by slicing a large training run into many
smaller ones that each stay under the cap (task splitting). This
underscores the need for coordinated verification across distributed
systems and possibly additional mechanisms (such as cluster size limits
or centralized license servers) to detect and prevent this form of
evasion. Verification may also include physical audits or on-site
inspections for high-risk actors, particularly those operating large
compute clusters.

While full-scale deployment of tamper-proof FLOP caps will take time,
especially if they rely on chip redesigns or new manufacturing steps,
their potential to reliably constrain compute makes them an important
long-term safeguard. Given the stakes, early investment in research and
prototyping is essential.

\hypertarget{offline-licensing-enforcing-compute-limits-through-time-limited-access-controls}{%
\subsubsection{Offline Licensing: Enforcing Compute Limits Through
Time-limited Access
Controls}\label{offline-licensing-enforcing-compute-limits-through-time-limited-access-controls}}

Offline licensing is a mechanism that limits how much computation a chip
can perform before requiring external authorization. By assigning chips
a finite compute budget (measured in FLOPs, and as discussed in the
preceding section) this licensing approach ensures that training runs
above a certain scale are only possible with renewed, externally
validated access.

\textbf{Governance} of such a mechanism
would require that a central authority (e.g. national regulator or
treaty body) would define FLOP quotas and issue usage or time-limited
licenses to hardware operators. Once a chip exhausts its budget, its
users must request a new license to continue. This governance model
gives key stakeholders the ability to approve or deny compute renewals
based on current risk levels, legal status, or compliance with AI safety
protocols. Critically, this can function even in air-gapped
environments, allowing global enforcement even without persistent
internet connectivity.

A real-world precedent is Intel's
``On Demand'' model for Xeon processors \parencite{intelIntelDemand}, where features can be remotely
activated via firmware. A similar paradigm could be repurposed for
licensing compute access, provided manufacturers include such
capabilities in future hardware generations, which would likely require
regulatory or treaty-based mandates.

\textbf{Enforcement} would occur through mechanisms embedded in firmware or hardware
that count FLOPs and enforce automatic shutdowns or severe slowdowns
once a chip's licensed budget is reached \parencite{petrieNearTermEnforcementAI2024a}. The license could be delivered
digitally or via physical media. Firmware-based approaches could be
implemented relatively quickly but may be vulnerable to tampering by
highly resourced actors. Hardware-level enforcement (with anti-tamper
protections) is more robust but will take longer to develop and deploy.

A key enforcement risk is license forgery: unauthorized parties might
attempt to generate fake licenses to extend compute use. This highlights
the need for secure key management and cryptographically verifiable
license issuance.

\textbf{Verification} hinges on whether the license renewal process is
controlled by a trusted authority. Chips would periodically ``check in''
--- digitally or through human-mediated workflows --- to request license
extensions. While approvals could be automatic under normal conditions,
the system would enable denial of renewals in high-risk contexts: e.g.
if a chip is moved to a prohibited jurisdiction, is linked to a
disallowed actor, or falls under new export control regimes. In this
way, offline licensing acts as a controllable pause button, enabling
governments or international bodies to halt compute-intensive AI
development dynamically, even after the hardware has already been deployed.

\hypertarget{fixed-set-mechanisms-preventing-large-compute-clusters-by-capping-ai-chip-networking-capabilities}{%
\subsubsection{\texorpdfstring{Fixed Set Mechanisms: Preventing
large compute clusters by capping AI chip networking capabilities
}{Fixed Set Mechanisms: Preventing large compute clusters by capping AI chip networking capabilities }}\label{fixed-set-mechanisms-preventing-large-compute-clusters-by-capping-ai-chip-networking-capabilities}}

One proposed method for limiting large-scale training runs is to
constrain how many AI chips can be effectively connected: either by
capping inter-chip bandwidth or fixing the number of GPUs that can
communicate during a training run. By preventing the formation of
high-throughput clusters, this ``fixed set'' approach seeks to cap the
compute that can be concentrated in any single training effort.
Fixed-set mechanisms may offer modest near-term benefits, especially
through fast-deployable infrastructure constraints.

To implement this from a \textbf{governance} perspective, international
or national regulators would need to mandate limits on interconnect
bandwidth or maximum cluster sizes for frontier-relevant hardware. Chip
manufacturers would then be required to produce hardware that enforces
these limits, either intrinsically (through chip design) or extrinsically
(through facility-level infrastructure constraints). Data centers would
need to be brought into compliance through certification and oversight
schemes.

In terms of \textbf{enforcement}, two main enforcement paths can be
pursued. The first is to enforce external infrastructure limits, such as requiring
data centers to use low-throughput cables \parencite[73486]{industryandsecuritybureauImplementationAdditionalExport2023} or to enforce physical
separation between chip ``pods'' \parencite{kulp2024hardware}. These methods are relatively fast to
deploy and do not require new hardware, but depend on facility-level
integrity and ongoing oversight. The second path is to exert intrinsic
chip-level controls, such as hard-coded limits on how many chips can
interconnect or bandwidth caps enforced via firmware or dedicated logic.
These are slower to implement (likely requiring multiple years of R\&D
and manufacturer cooperation) but are more robust over time.

However, both approaches face growing feasibility challenges. Recent research from
DeepMind demonstrates that large-scale training can now be distributed
with two orders of magnitude less interconnect bandwidth than was
previously required, while achieving comparable model performance \parencite{douillardStreamingDiLoCoOverlapping2025}. This undermines
the core premise that limiting interconnect bandwidth will significantly
delay or degrade frontier model training \parencite{clarkImportAI3982025}. As distributed training
becomes more efficient, the effectiveness of fixed-set mechanisms may
rapidly diminish.

\textbf{Verification} would require ensuring that physical and logical
interconnect limits are actually enforced. This could involve active
monitoring, such as in-person audits, surveillance cameras, and cable
inspections, tamper-resistant enclosures, to prevent reconfiguration of
chip clusters after inspection, and intrinsic telemetry or receipts,
showing interconnect topology and chip communication logs.

However, verification also faces evasion risks. For example, a training
run could be split into many smaller sub-tasks that individually obey
cluster limits but collectively perform a threshold-exceeding
computation. Without mechanisms to link and audit compute usage across
time or across disconnected clusters, such circumvention may be
difficult to detect.

While fixed-set mechanisms may offer some near-term benefits, their
long-term utility is increasingly questionable due to rapid progress in
distributed training methods and the difficulty of reliably detecting
task fragmentation. As such, they should be pursued only in combination
with other mechanisms that directly track or limit total compute usage
over time.

\hypertarget{model-locking-preventing-the-unauthorized-deployment-of-model-weights-outside-of-a-secured-environment}{%
\subsubsection{Model Locking: Preventing the unauthorized
deployment of model weights outside of a secured
environment}\label{model-locking-preventing-the-unauthorized-deployment-of-model-weights-outside-of-a-secured-environment}}

Model locking is a mechanism designed to prevent the unauthorized use or
distribution of AI model weights after training. It allows researchers
to conduct
training and evaluation, but ensures that deployment or transfer of the
model requires explicit approval from a third party \parencite{petrie2024interim}. Model locking
offers a useful governance lever for slowing deployment and requiring
centralized approval for release. It enhances control over how and when
powerful models are used, but does not prevent internal misuse
post-training. As such, it should be paired with other mechanisms (such
as FLOP caps or usage-based licensing) that constrain risky model
development before weights are produced.

\textbf{Governance} of this approach requires AI chips to encrypt model
weights during training, such that the resulting models cannot be
accessed or deployed without a decryption key. Only authorized chips (or
chips held by a trusted oversight body) would be capable of decrypting
and running the trained model. Governance, in this case, involves a
centralized or multilateral entity controlling the issuance of
decryption keys. This oversight body could be a national regulator,
treaty organization, or consortium of safety evaluators who assess the
risks of a model before approving its release.

Chip manufacturers would need to build secure processors that act as
intermediaries controlling the input and output for a GPU. This
architecture could require that any transfer of weight shards off of the
chip are encrypted by default to block unapproved transfers. Researchers
using these chips could train and test their models, but the model
weights would remain unusable outside of authorized environments unless
the oversight body grants a key. \textbf{Enforcement} would require
policy mandates that restrict large-scale or high-risk training to chips
equipped with this locking mechanism.

Importantly, this mechanism limits external misuse and unauthorized
public release, for instance, by deterring model theft or premature mass
deployment. However, it does not prevent misuse by the training lab
itself, if the model is executed internally for harmful or misaligned
purposes. Therefore, model locking is best understood as a tool to delay
or gate deployment, not as a full safeguard against all forms of loss of
control.

\textbf{Verification} could include audits of training facilities to
confirm that model-locking hardware is in use for high-risk projects.
Additionally, if a model trained under a locking regime is released
without authorization, it would constitute a clear breach. Monitoring
public model releases by known labs could help detect circumvention.
More advanced governance schemes could involve multi-party decryption
protocols (e.g. requiring several parties to jointly authorize release),
increasing robustness in multipolar governance scenarios.

\hypertarget{traceability-intervention-point}{%
\subsection{Traceability Intervention
Point}\label{traceability-intervention-point}}

In this subsection and the next, we discuss traceability and regulatory
intervention points. While we believe there is a meaningful conceptual
distinction between them, they are closely intertwined, and it is worth
clarifying how to approach them in this framework.

Traceability mechanisms generate visibility into compute flows: they
help us observe and track where compute goes, who uses it, and how.
Regulatory mechanisms, by contrast, establish the legal authority to
constrain and act on that visibility: they define who gets to do what,
and under what rules.

It may be tempting to treat traceability as purely about verification,
and regulation as purely about governance and enforcement. But that
would be a mistake. You can have traceability without regulatory force
(e.g., voluntary disclosures), and regulation without traceability
(e.g., laws on paper with no means of oversight). But both are
inadequate on their own.

As we noted earlier, the GEV framework operates like a tripod: remove
one leg --- governance, enforcement, or verification --- and the entire
system risks collapse. Importantly, this applies not just to the overall
pause architecture, but to each intervention point within it.
Traceability mechanisms require clear governance (e.g., which flows must
be reported), robust enforcement (e.g., penalties for failing to
disclose), and credible verification (e.g., audits or automated
reporting). Regulatory tools likewise must be paired with mechanisms for
monitoring compliance and verifying outcomes. Each layer of intervention
must be internally sound across GEV dimensions for the overall system to
function.

The traceability intervention point aims to make the compute supply
chain and access ecosystem fully observable: from the sourcing of
critical manufacturing equipment, to the volume and location of chip
production, to the movement and use of compute resources by end users.
These mechanisms do not impose restrictions directly, but instead
provide the data infrastructure and transparency needed to detect
violations, inform enforcement, and support global coordination.
Effective traceability enables policymakers to monitor who is producing
chips, where they are going, and how they are being used, thereby
forming the informational backbone of the Compute Pause Button.

\begin{table}[ht]
\centering
\caption{Traceability Mechanisms Overview and Distinctions}
\label{tab:traceability_mechanisms}
\begin{tabularx}{\linewidth}{@{} 
  >{\raggedright\arraybackslash}X 
  >{\raggedright\arraybackslash}X 
  >{\raggedright\arraybackslash}X 
  @{}}
\toprule
\textbf{Mechanism} & \textbf{Primary object of traceability} & \textbf{Primary purpose} \\
\midrule
Monitoring Key Materials and Chip Components &
Suppliers and manufacturers of critical chip-making equipment and components &
Prevent illicit chip production by monitoring the flow of essential manufacturing inputs \\
\midrule
Global Compute Supply Registry &
Chipmakers and fabrication facilities producing high-performance AI chips: production volumes, technical specifications, and sales records &
Enable global visibility into chip output and distribution patterns for policy oversight (including production and export controls) \\
\midrule
Chain-of-Custody for High-Performance Chips &
Entities and facilities in possession of high-performance chips after manufacture: movement, ownership, and declared use &
Detect unauthorized deployment or stockpiling of chips after manufacture \\
\midrule
Know Your Customer (KYC) &
Users requesting access to large-scale cloud compute resources: identity, intent, and compute usage patterns &
Ensure that access to large-scale compute is limited to verified, approved users \\
\bottomrule
\end{tabularx}
\end{table}

\hypertarget{monitoring-key-materials-and-chip-components-preventing-illicit-compute-production}{%
\subsubsection{Monitoring Key Materials and Chip Components:
Preventing illicit compute
production}\label{monitoring-key-materials-and-chip-components-preventing-illicit-compute-production}}

To prevent the unauthorized production of high-performance AI chips, a
traceability regime could be established to monitor the movement of
critical manufacturing tools and components across the global
semiconductor supply chain. By tracking these upstream chokepoints ---
which are highly concentrated, difficult to replicate, and necessary for
frontier manufacturing --- regulators can reduce the risk of covert chip
production outside of approved channels.

An international oversight body (potentially modeled after the IAEA
which tracks nuclear materials under the Non-Proliferation
Treaty or the Organization for the Prohibition of Chemical Weapons
(OPCW)) would coordinate declarations, inspections, and reporting
protocols for firms involved in chip tool manufacturing, component
production, and advanced fab operations. It would coordinate
declarations, audit supply chains, and manage a secure registry of
regulated items. This body could be constituted either as a formal
treaty organization or a coordinated regime among major chip-producing
nations.

\textbf{Governance} would begin with the creation of a register of
regulated components considered critical for frontier chip production.
This list must be updated regularly to reflect technological advances.
It could include, among others:

\textbf{Tier 1: Core Manufacturing \& Design Hardware:}

\begin{itemize}
\item
  Extreme Ultraviolet EUV and \href{https://www.zeiss.com/semiconductor-manufacturing-technology/inspiring-technology/high-na-euv-lithography.html}{future High-NA EUV lithography systems} \parencite{khanSecuringSemiconductorSupply2021}.
\item
  Advanced electronic design automation (EDA) software and associated IP cores.
\item
  Advanced ion implantation and wafer annealing equipment.
\item
  Metrology and inspection tools from specialized suppliers (e.g., KLA, Applied Materials) \parencite{bureauofindustryandsecurityCommerceStrengthensExport2024}.
\item
  Key optics and laser systems (e.g., Zeiss SMT lenses, Cymer lasers).
\end{itemize}

\textbf{Tier 2: Critical Components \& Consumables:}

\begin{itemize}
\item
  EUV photomasks (reticles), blanks, and pellicles.
\item
  Ultra-pure process gases (e.g., tungsten hexafluoride, disilane) and specialty photoresists.
\item
  High-purity quartz and silicon carbide components for process chambers.
\end{itemize}

\textbf{Tier 3: Advanced Packaging Equipment:}

\begin{itemize}
\item
  Wafer-to-wafer and die-to-wafer hybrid bonding machines.
\end{itemize}

\begin{itemize}
\item
  \parbox{\linewidth}{%
    Equipment for producing High-Bandwidth Memory (HBM) and implementing 
    System-in-Package (SiP) designs like TSMC's CoWoS.%
  }
\end{itemize}

\textbf{Enforcement} would rely on mandatory declarations, tamper-resistant transfer records, and coordination with customs and border authorities to prevent diversion. If violations occur (such as undeclared transfers or mismatched inventories) the regulatory body could trigger a range of penalties, including denial of future access to regulated components, export controls on offending firms, and coordinated international sanctions. This structure mirrors enforcement protocols under the \protect\hyperlink{policy-analogue-2-the-chemical-weapons-convention-cwc-and-the-organisation-for-the-prohibition-of-chemical-weapons-opcw}{Organization for the Prohibition of Chemical Weapons} (OPCW), where inspection-triggered violations lead to sanctions and removal of access to controlled substances.

\textbf{Verification} would be implemented through a mix of chain-of-custody tracking (e.g., barcodes, QR codes, or RFID tags for physical parts), supply chain audits and documentation, on-site inspections of suppliers and fabs, and due diligence obligations for downstream firms. The traceability of "3TG" conflict minerals (i.e., tin, tungsten, tantalum, and gold) under the U.S. Dodd-Frank Act provides a useful precedent, combining documentation, third-party audits, and centralized registries to enforce transparency in complex global supply chains. These measures would create multiple layers of verification designed to detect and prevent unauthorized diversion of critical AI chip equipment and components before they could be used for illicit manufacturing.

\hypertarget{global-compute-supply-registry-requiring-all-chip-production-facilities-to-register-with-a-central-verification-body-to-log-volumes}{%
\subsubsection{Global Compute Supply Registry: Requiring all chip
production facilities to register with a central verification body to
log
volumes}\label{global-compute-supply-registry-requiring-all-chip-production-facilities-to-register-with-a-central-verification-body-to-log-volumes}}

To prevent the unchecked spread of high-performance AI chips, a global
compute supply registry would require chip manufacturers, fabs, and
semiconductor suppliers to report production volumes, technical
specifications, and sales records to an internationally coordinated
body. This registry would act as a transparency and risk assessment
tool, enabling policymakers and regulators to detect suspicious
activity, assess global supply trends, and close gaps in enforcement
mechanisms such as export controls or pause regimes. Rather than
directly restricting trade, the registry would support a broader compute
governance architecture by providing real-time visibility into who is
producing frontier chips, in what quantities, and where they are going.
Analogous to the Nuclear
Suppliers Group or IAEA
safeguards, this transparency mechanism helps ensure that
high-performance compute capacity does not quietly proliferate beyond
agreed safety thresholds.

The first step in \textbf{governance} is building a global registry is
to define
what qualifies as a high-performance AI chip \parencite{industryandsecuritybureauFrameworkArtificialIntelligence2025}. This would require
standardized technical benchmarks, such as those offered by MLPerf
\parencite{mlcommonsBenchmarkMLPerfTraining}, which evaluate hardware performance for AI
inference and training tasks. A multilateral working group, comprising
technical agencies like the National
Institute of Standards and Technology (NIST), trade bodies such as the
EU Directorate-General for Trade and Economic Security, and major
chip-producing countries in the Chip
4 Alliance (U.S., Taiwan, South Korea, Japan), could formalize these
thresholds and coordinate implementation.

All manufacturers of qualifying chips would be required to register
their facilities and report production volumes and destinations. This
registry would be maintained by an international technical oversight
body, which could operate similarly to the Nuclear Suppliers Group or Wassenaar Arrangement,
providing shared visibility and signaling potential risks to national
regulators.

While the registry itself would not control exports, it would feed into
risk flags, compliance checks, and sanctions regimes that make up
\textbf{enforcement}. Governments could use registry data to deny
licenses to exporters with suspicious reporting patterns, investigate
diversion risks from downstream clients, and coordinate trade
restrictions against violators. Enforcement would be supported by a
combination of licensing regimes, supply chain contracts, and
chokepoints across the chip ecosystem, including foundries, EDA software
providers, and cloud platforms. Precedents include U.S. Export Administration
Regulations (EAR) regulations and recent Dutch-U.S.
restrictions on ASML EUV sales to China \parencite{shivakumar2024balancing}.

Regional coalitions (e.g., ASEAN's Compute South) could reinforce
enforcement by coordinating inspections, verifying local reporting, and
aligning penalties. Sanctions for noncompliance could include export
privilege revocation, market access bans, or substantial fines, thereby
creating strong deterrents against misreporting or non-disclosure.

\textbf{Verification} would involve regular and surprise audits of chip
production and distribution records. Inspectors, who would be appointed
by an international oversight body or national regulators, would
cross-check registry entries with fab output logs and facility-level
records, customs and shipping documents, and purchase and deployment
data from major cloud or compute users. To enhance traceability, the
system could incorporate digital tools such as blockchain-based chip
serialization, ensuring that registered chips are verifiably tracked
across jurisdictions. This mirrors best practices from nuclear
inspections and consumer tech regulation, such as Indonesia's
enforcement against non-compliant smartphone imports \parencite{reutersIndonesiaBansSales2024}.

Industry organizations like Semiconductor Equipment and
Materials International (SEMI) could support the development of
standardized reporting protocols and data formats to ensure global
interoperability. Together, these tools would create a scalable,
tamper-resistant audit trail that spans from chip fabrication to
deployment.

\hypertarget{know-your-customer-kyc-schemes-vetting-access-to-large-scale-compute-resources}{%
\subsubsection{Know Your Customer (KYC) Schemes: Vetting access to
large-scale compute
resources}\label{know-your-customer-kyc-schemes-vetting-access-to-large-scale-compute-resources}}

Know Your Customer (KYC) is a regulatory practice used to verify the
identity and intent of clients before granting access to sensitive
services or resources. Originally
developed to combat financial crimes like money laundering and
terrorist financing \parencite{swiftKnowYourCustomer}, KYC procedures are now being proposed for
large-scale cloud compute providers to help govern access to AI training
infrastructure. The core idea is to ensure that those requesting
significant computational resources are verified, accountable, and
operating within approved boundaries. While KYC is a regulatory tool in
origin, we categorize it under traceability here because its primary
function in the compute governance context is to generate visibility and
auditability over user access, thereby making it a critical complement
to hardware- and supply-side tracing efforts.

Applying KYC to AI compute is a natural extension of its success in
finance. By tying compute access to verified identities, declared
project purposes, and auditable records, this approach creates a
transparent, traceable system that deters misuse. It ensures that
high-risk actors (such as sanctioned entities or anonymous users seeking
to train powerful models) are blocked or flagged. The U.S. government
has already signaled its intent to apply KYC in this domain, with the
Department of Commerce proposing rules under Executive
Order 13984 to require cloud providers to vet foreign users in light
of cyber risks \citeyear{TakingAdditionalSteps2021}. This sets a strong precedent for extending such systems
to monitor and enforce AI governance rules globally.

\textbf{Governance} would require governments or an international
regulatory body to mandate that large-scale compute providers (e.g. AWS,
Microsoft Azure, Google Cloud) implement KYC protocols for users
exceeding defined compute thresholds. This would include identity
verification, documentation of intended AI projects, and confirmation of
compliance with safety regulations. Cloud providers would be legally
obligated to collect and maintain these records and report suspicious
activity to oversight bodies.

This system could be tied to broader compute licensing regimes or pause
mechanisms: only verified and approved users could access the levels of
compute needed to train frontier AI systems. Governance mandates could
also require that KYC systems align with global sanction lists like the Office of Foreign Assets Control (OFAC)
or the UN Security Council Consolidated List to block access by flagged or
high-risk actors.

\textbf{Enforcement} would operate through mandatory identity-linked
logging of compute access, along with monitoring and reporting systems.
Cloud providers would log all access and usage patterns, tied to unique
user identities and then be required to deny compute access above
regulatory thresholds to unverified users. They would also report
suspicious activity or unauthorized requests to regulators. Penalties
for noncompliance could escalate from fines to access bans or even
criminal prosecution. As an illustrative example, the U.S.
Treasury fined TD Bank \$1.3 billion in 2024 for systemic failures in
implementing KYC protocols \parencite{financialcrimesenforcementnetworkFinCENAssessesRecord2024}, demonstrating the seriousness with which
enforcement can be pursued.

\textbf{Verification} would combine technical monitoring, regular
audits, and automated cross-checks. Regulators would inspect cloud
providers' KYC records, usage logs, and system configurations to ensure
proper implementation. Identity checks would be cross-referenced with
global watchlists to flag unauthorized users. This system could be
further strengthened by integrating chip-level reporting mechanisms (as
proposed in
\protect\hyperlink{technical-intervention-point}{other sections
of the pause button framework}) that automatically log large-scale
compute usage and transmit metadata to
oversight entities \parencite{heimGoverningCloudIntermediary2024}. Together, these verification steps ensure that
access to powerful compute is traceable, auditable, and revocable.

\hypertarget{chain-of-custody-framework-for-high-performance-chips-tracking-the-ownership-and-movement-of-finished-chips-to-prevent-diversion-or-misuse}{%
\subsubsection{Chain-of-Custody Framework for High-Performance Chips
→ Tracking the ownership and movement of finished chips to prevent
diversion or
misuse}\label{chain-of-custody-framework-for-high-performance-chips-tracking-the-ownership-and-movement-of-finished-chips-to-prevent-diversion-or-misuse}}

A chain-of-custody system for high-performance AI chips would establish
a verifiable record of where each chip goes, who controls it, and how it
is used. Similar to systems used to track 
\href{https://worldbigroup.com/Event-blogs/track-&-trace-systems#:~:text=The%20RFID%20system%20consists%20of,in%20meeting%20those%20mandatory%20requirements.}{pharmaceuticals}, 
\href{https://www.nicepublicsafety.com/glossary/evidence-management-software}{evidence in legal cases}, or 
\href{https://landairsea.com/blogs/consumers/what-is-satellite-tracking}{satellite assets}, this approach aims to prevent unauthorized deployment, illicit
resale, or covert concentration of compute. The system would begin with
unique chip registration at the point of manufacture and continue
through a tamper-resistant, centralized database that logs serial
numbers, ownership history, declared end-use, and any transfers across
organizations or borders.

This intervention is especially valuable in the context of a global
pause or compute cap: it enables regulators to detect hidden compute
clusters, restrict access to powerful hardware, and identify violations
in near real-time. A useful parallel is Intelsat,
which tracks and audits satellite ownership and deployment globally,
even without being a formal regulatory authority \parencite{macaskill2025intelsatasamodel}. However, AI chips pose
unique challenges: millions are produced annually, they are easily
portable, and they power a wide range of civilian and commercial uses.
As such, this mechanism must be automated, scalable, and globally
coordinated to be effective.

\textbf{Governance} for a global chain-of-custody regime would require
all high-performance AI chips (e.g., GPUs, TPUs, ASICs above a compute
threshold) to be registered at the point of manufacture. Each chip would
receive a cryptographic identifier --- either embedded in hardware or
linked to firmware --- that allows it to be tracked through its
lifecycle. A centralized, tamper-proof registry would store data on chip
ownership, transfers, usage location, and declared purposes.

Governance responsibilities could be handled by a newly created
international technical oversight body, modeled on organizations like
the IAEA or Intelsat.
While it may not hold formal enforcement authority over sovereign
nations, such a body could set standards, maintain the registry, and
issue reputational or technical flags that inform the actions of
national regulators and suppliers. It could also facilitate
interoperability between national customs databases, cloud providers,
and chip vendors.

A strong \textbf{enforcement} regime would ensure that unregistered or
misused chips cannot operate undetected. National customs agencies would
inspect chip shipments, verify registration records, and seize
undeclared or blacklisted hardware. Domestically, access to sensitive
inputs (such as firmware updates, design software, cloud infrastructure,
and even electrical or internet connectivity) could be restricted for
non-compliant actors. In practice, enforcement would rely heavily on
interdependence within the semiconductor ecosystem. Governments could
pressure violators by coordinating export controls, cutting off access
to critical tools like ASML lithography machines or EDA software, or
denying upstream support from vendors. This form of soft enforcement
through ecosystem leverage is especially useful when direct legal
authority is fragmented across borders.

\textbf{Verification} ensures that the registry is accurate and that
declared usage matches actual deployment. Facilities hosting high-risk
chips (such as AI labs or large datacenters) would undergo routine
audits by certified third parties. These audits could review compute
utilization logs, chip usage declarations, facility access records,
physical security measures (e.g., tamper-proof enclosures, access
control systems), and cybersecurity protections (e.g., firmware
integrity checks, network segmentation). In addition, automated anomaly
detection systems could flag unexpected behavior, for example, if a chip
registered to one lab is suddenly reporting usage from a different
region, or if compute hours exceed declared project requirements.
Facilities might also be subject to surprise inspections to check for
hidden or diverted chips.

To account for global scalability, particularly in countries with
limited regulatory capacity, international funding or technical support
may be needed to build out verification infrastructure. Without routine
and credible verification, chain-of-custody systems risk becoming
symbolic rather than operational. While the logistics involved in
chain-of-custody monitoring are non-trivial, the combination of unique
chip identifiers, centralized registries, and ecosystem enforcement
leverage makes this a feasible and impactful intervention, particularly
when combined with facility audits and automated usage monitoring.

\hypertarget{regulatory-intervention-point}{%
\subsection{Regulatory Intervention
Point}\label{regulatory-intervention-point}}

The regulatory intervention point encompasses mechanisms that define,
formalize, and enforce the rules governing who can access, produce, or
export high-performance compute. These tools do not monitor or track
activity directly; instead, they create the legal and institutional
authority to constrain it. Regulatory mechanisms such as export
controls, production caps, and licensing regimes give national and
international actors the power to impose hard limits on the scale,
spread, and use of advanced AI chips. While traceability mechanisms
provide the visibility needed to detect violations, regulation provides
the teeth to act on them, thereby determining which actors are
authorized, what thresholds are acceptable, and what consequences follow
from noncompliance.

\begin{table}[htbp]
\centering
\caption{Regulatory Mechanisms Overview and Distinctions}
\label{tab:regulatory_mechanisms}
\begin{tabularx}{\linewidth}{@{} 
  >{\raggedright\arraybackslash}X 
  >{\raggedright\arraybackslash}X 
  >{\raggedright\arraybackslash}X 
  @{}}
\toprule
\textbf{Mechanism} & \textbf{Primary object of traceability} & \textbf{Primary purpose} \\
\midrule
Production Controls &
Chipmakers producing high-performance AI &
Limit the total volume of high-performance chips produced to slow uncontrolled compute scaling \\
\midrule
Export Controls &
Entities purchasing or receiving high-performance AI chips across borders &
Prevent dangerous actors or jurisdictions from acquiring frontier compute through targeted trade \\
\bottomrule
\end{tabularx}
\end{table}

\hypertarget{production-controls-limiting-the-volume-and-type-of-high-performance-chips-via-quotas-and-licensing}{%
\subsubsection{Production Controls: Limiting the Volume and Type of
High-Performance Chips via Quotas and
Licensing}\label{production-controls-limiting-the-volume-and-type-of-high-performance-chips-via-quotas-and-licensing}}

Production controls aim to slow the unchecked scaling of frontier AI
systems by limiting how many high-performance chips (such as GPUs, TPUs,
and ASICs) can be manufactured globally. This mechanism would introduce
ceilings on chip output, tied to safety, security, and strategic
objectives. By capping chip production through licenses and global
coordination, this approach reduces the hardware available for
unmonitored large-scale training runs and creates a policy lever for
pacing AI progress. It also gives regulators time to implement safety
evaluations and downstream controls without falling behind the
exponential curve of compute growth.

\textbf{Governance} would begin with national and international
authorities setting clear thresholds for what qualifies as
``high-performance AI chips'' (based on compute density, memory
bandwidth, or ML benchmark performance, e.g. MLPerf). Countries or
blocs (e.g., the U.S., EU, Taiwan, South Korea, Japan) would allocate
production quotas to domestic manufacturers, coordinated through an
international agreement or Chip Producers Alliance.

This model draws inspiration from The U.S.
Drug Enforcement Administration, which allocates annual production
quotas for controlled substances based on projected legitimate use \parencite{drugenforcementadministrationProposedAggregateProduction2024},
and from OPEC's oil production
coordination, which balances output with strategic and economic
priorities. Governments could adjust chip quotas over time in response
to risk assessments, demand trends, geopolitical stability, or hardware
stockpiles, giving them a flexible lever to stabilize supply and
mitigate proliferation risk.

\textbf{Enforcement} of production limits would be implemented through a
licensing system, requiring chipmakers to obtain approval to manufacture
above defined thresholds. National regulatory bodies would oversee
license issuance and ensure that manufacturers comply with reporting and
quota obligations. Sanctions for noncompliance could include suspension
or revocation of production licenses, fines or civil penalties for
exceeding quotas or falsifying data, blacklist status preventing access
to critical upstream inputs (e.g., photolithography tools, rare earth
elements, or design software), and exclusion from public contracts or
access to regulated markets.

A relevant historical parallel is the U.S.
Agricultural Adjustment Act of 1933, where the government incentivized
farmers to reduce crop production through subsidies and penalties \parencite{libraryofcongressUnitedStatesCode}. A
similar mix of incentives, licenses, and sanctions could be adapted for
high-performance AI chips to ensure semiconductor manufacturers stay
within global safety limits.

\textbf{Verification} would involve a combination of physical audits and
secure digital production tracking. Chipmakers would log each wafer
batch onto a tamper-proof digital ledger (potentially blockchain-based)
with unique identifiers and metadata. National regulators, supported by
international inspection teams, would verify these records against
factory output and shipping data.

Drawing on practices from pharmaceutical regulation, agencies like Food and Drug Administration (FDA)
and the European Medicines Agency (EMA) could provide useful models for \href{https://www.who.int/teams/health-product-policy-and-standards/standards-and-specifications/norms-and-standards/gmp}{Good Manufacturing Practice (GMP)}. These bodies conduct rigorous audits of
production facilities to assess hygiene, equipment calibration, and
documentation. Applied to chip production, similar protocols would
assess actual vs. reported output, compliance with licensing terms, and
security of facilities and digital systems. Violations could trigger
export bans, facility shutdowns, or disconnection from global supply
chains. Real-time monitoring, third-party audits, and random inspections
would reinforce the credibility of the system.

\hypertarget{export-controls-regulating-the-specific-actors-and-countries-that-can-access-which-chips-and-how-many}{%
\subsubsection{Export Controls: Regulating the specific actors and
countries that can access which chips and how
many}\label{export-controls-regulating-the-specific-actors-and-countries-that-can-access-which-chips-and-how-many}}

Export controls are a well-established tool for managing the spread of
strategically sensitive technologies. Applied to frontier AI, a global
export control regime would restrict the sale and transfer of
high-performance AI chips to specific actors and destinations, in
quantities that reflect safety, geopolitical, and developmental
considerations. This mechanism is essential to ensure that the raw
compute power required for training dangerous AI systems is not acquired
covertly or stockpiled by unregulated actors. Rather than banning chip
sales entirely, this system would function through targeted control
lists, pre-approval requirements, and buyer-specific quotas, thereby
ensuring that only trusted actors can purchase powerful chips, and only
in regulated quantities.

In terms of \textbf{governance}, a multilateral governance body --- such
as a Global AI Oversight Council convened under the G7, OECD, or UN ---
would define the parameters of the export control regime. Its
responsibilities would include setting technical thresholds for
regulated chips (e.g., FLOPs, memory bandwidth, interconnect speed),
maintaining control lists of chip types subject to export approval,
developing buyer classification systems to determine who may receive
chips and in what quantity, and coordinating regular updates to reflect
rapid changes in chip capabilities and supply chains.

This framework draws directly from the Wassenaar
Arrangement, which governs dual-use export controls for military
and sensitive technologies \parencite{hrynkiv2024not}. Under Wassenaar, states agree on shared
control lists and commit to implementing export restrictions at the
national level. A similar structure for AI chips would allow individual
countries to retain enforcement authority while aligning with global
safety standards.

\textbf{Enforcement} would take place through national export control
agencies, which would issue (or deny) export licenses based on the
technical specifications of the chips, the identity and classification
of the buyer, the intended end-use, and the volume of chips requested.
To ensure meaningful deterrence, enforcement measures could deny export
licenses to flagged entities and apply quotas or ceilings on how many
chips a buyer can legally purchase. Sanctions for violations could
involve substantial financial penalties, suspension of export
privileges, and loss of access to upstream vendors (e.g., EDA tools or
chip firmware updates). Precedents include the EU's
Digital Markets Act \parencite{kissBrusselsEffectHow2024}, which has shown that well-resourced governments
can compel compliance from even the largest tech firms through
structured legal regimes.

\textbf{Verification} mechanisms would ensure that chips are not
diverted after approval or used by unauthorized parties. These would
include end-use and end-user reporting requirements, on-site inspections
of major buyers or facilities of concern, secure chip-level identifiers
to track whether exported chips end up at their declared destinations,
and information sharing between participating states (as practiced under
Wassenaar), especially regarding denied licenses or suspicious requests.
Any discrepancy between declared and actual chip deployment (such as
unauthorized resale, failure to report usage, or excessive stockpiling)
would trigger penalties and additional scrutiny.

This approach could be supplemented by supply chain audits and chip
registry cross-checks, creating redundancy in the verification system.
Unlike mechanisms focused on internal governance (e.g., KYC or
licensing), export controls operate at the border, making them one of
the few tools capable of preemptively stopping dangerous compute
accumulation.

\hypertarget{relevant-historical-policy-analogues-precedents-for-global-coordination-and-the-need-for-institutional-imagination}{%
\section{Relevant Historical Policy Analogues: Precedents for Global
Coordination and the Need for Institutional
Imagination}\label{relevant-historical-policy-analogues-precedents-for-global-coordination-and-the-need-for-institutional-imagination}}

In designing a credible and enforceable mechanism to pause frontier AI
training runs, we need more than technical tools: we need
\textbf{institutional imagination}. While no existing regime governs AI
compute in the way we propose in this paper, history offers several
analogues that grapple with similar challenges, i.e., controlling
dangerous dual-use technologies, coordinating across jurisdictions,
enabling robust verification, and sustaining compliance among powerful
actors. This section explores four such policy arrangements, each
offering instructive insights into how global coordination and oversight
can function in high-stakes domains, despite some foreseeable
shortcomings.

Our approach draws on familiar elements from global governance design.
Most closely, it echoes enforcement-driven institutions, which pair
binding treaties with robust inspection and verification powers. At the
same time, it incorporates aspects of policy coordination seen in
regimes like the World Trade
Organization (WTO) or Financial Action
Task Force (FATF) (such as registries, certification systems, and
harmonized national implementation). In early phases, soft-law features
like voluntary KYC protocols or compute transparency initiatives may
also mirror norm-setting bodies that build alignment before formal
treaties emerge.

Specifically, in this paper we examine: the NPT/IAEA system's binding
verification infrastructure for nuclear technologies; the CWC/OPCW
regime's handling of precursor chemical materials and hidden threats;
Operation Warp Speed as a model of agile, mission-driven governance
under urgency; and the Wassenaar Arrangement's soft-law export control
framework for dual-use technologies. While these examples vary in scope,
coerciveness, and structure, they provide scaffolding for imagining how
a global compute pause mechanism might be governed, enforced, and
verified. To better analyze their relevance, we group them into two
categories: binding regimes with robust legal and verification
frameworks, and non-binding coordination efforts that leverage norms,
incentives, and organizational innovation. Both offer vital lessons for
designing effective AI governance, but from different ends of the
spectrum.

While some of the policy analogues in this section rely on soft law ---
i.e., encouraging cooperation or compliance by using transparency,
coordination, or market incentives --- our proposed mechanisms stress
the need for a binding, enforceable, and verifiable approach. Voluntary
measures will not suffice given that frontier AI has potentially
devastating consequences and when the incentives (both economic and
political) to defect from soft agreements are too strong. Still, soft
law examples offer useful lessons for early coordination, industry
signaling, and pathways toward more robust governance treatises and
institutions.

\hypertarget{binding-regimes-hard-law-hard-verification}{%
\subsection{Binding Regimes: Hard Law, Hard
Verification}\label{binding-regimes-hard-law-hard-verification}}

The first two policy cases --- the NPT/IAEA and the CWC/OPCW --- offer
the strongest precedents for AI governance mechanisms that demand
enforceable commitments, technical inspections, and international
oversight. They represent what might be called the ``hard law''
approach: treaty-based systems with legal obligations, institutional
authority, and verification architectures.

\hypertarget{policy-analogue-1-the-nuclear-non-proliferation-treaty-npt-and-the-international-atomic-energy-agency-iaea}{%
\subsection{Policy Analogue 1: The Nuclear Non-Proliferation
Treaty (NPT) and the International Atomic Energy Agency
(IAEA)}\label{policy-analogue-1-the-nuclear-non-proliferation-treaty-npt-and-the-international-atomic-energy-agency-iaea}}

\textbf{Analogue Snapshot: Controlling catastrophic dual-use technology
through international governance, monitoring, and enforcement}

Few regimes have faced higher stakes than the one tasked with preventing
the spread of nuclear weapons. The NPT,
backed by the IAEA, is a landmark
effort to contain nuclear energy, which is a powerful dual-use
technology, for which peaceful applications exist alongside devastating
military potential. It's also one of the clearest historical analogues
for the governance of advanced AI: a potentially highly dangerous, yet
positively transformative technology with global consequences.

The NPT sets the political and legal commitments among nations to
prevent nuclear weapons proliferation, while the IAEA provides the
technical infrastructure and independent verification needed to monitor
compliance; together, they create a system where rules are both agreed
upon, enforced, and verified.

\textbf{GEV Mapping}

\textbf{Governance}

The NPT establishes binding international commitments. Countries join
voluntarily but must accept clear restrictions on weaponization in
exchange for access to peaceful nuclear technology, such as nuclear
energy generation, medical isotope production, and agricultural
applications like food irradiation. In total, 191 states have joined the
Treaty, including the five officially-recognized nuclear-weapon states.
More countries have ratified the NPT than any other arms limitation and
disarmament agreement.

The IAEA serves as a technical body that interprets, implements, and
helps enforce these commitments. The rules are global, but
implementation is flexible and coordinated through national governments.
Analogously, our proposal would involve a global oversight body to
implement and coordinate mechanisms like production quotas, export
controls, and a global compute supply registry, defining limits on chip
manufacturing and circulation.

\textbf{Enforcement}

Violations of the NPT can lead to the loss of access to peaceful nuclear
technology, diplomatic isolation, and escalation to the UN Security
Council, which may impose sanctions or mandate other international
responses. Enforcement is collective: no single state acts alone, but
violations trigger coordinated responses, primarily through the IAEA's
reporting and the Security Council's authority.

A prominent test case is North Korea, which announced its withdrawal
from the treaty in 2003 under Article X of the NPT, citing
``extraordinary events'' threatening its national interests. However,
its withdrawal has never been universally recognized, and debate
continues over its legal status under the treaty. Some
analysts argue North Korea keeps its potential reentry into the NPT as
a diplomatic bargaining chip, leveraging the treaty's political
legitimacy even as it violates its core tenets \parencite{carrel2010nuclear}. This case highlights
both the challenges of enforcing previously agreed-upon rules and
penalties against noncompliant actors, and the residual power of
international agreements (even when violated) to shape long-term
behavior.

In the AI context, enforcement could include revoking offline compute
licenses, triggering chip self-deactivation via FLOP cap enforcement, or
imposing trade restrictions on actors found violating compute thresholds
or circumventing chain-of-custody protocols.

\textbf{Verification}

The IAEA runs one of the most advanced international monitoring systems
in existence, signed by member states. It uses a multilayered approach
that includes: on-site inspections at declared facilities (where
inspectors can access sensitive areas and equipment), remote
surveillance systems (including cameras and motion detectors, to ensure
continuous monitoring), environmental sampling (which can detect traces
of nuclear material and activities even if undeclared), satellite
imagery and open-source intelligence (to detect undeclared facilities),
and material accountancy systems (that track nuclear materials from
production through storage and use). This multilayered approach makes
noncompliance detectable in practice; it has repeatedly detected
violations and triggered international responses, such as uncovering Iran's
undeclared enrichment facilities and plutonium production in the early
2000s \parencite{internationalatomicenergyagencyVerificationMonitoringIran2014}, and identifying discrepancies in
North Korea's
nuclear declarations in the 1990s that led to its withdrawal from the
treaty and international sanctions \parencite{internationalatomicenergyagencyVerificationDPRK2014}.

The system shows how a technically sophisticated, politically backed
verification mechanism can provide credible, timely, and independent
verification of compliance, thereby making it a vital role model for AI
compute governance, where secrecy and scale can otherwise enable runaway
development.

Our proposed system mirrors this with a blend of KYC protocols for cloud
access, tracking of critical components, tamper-resistant chip
telemetry, and physical audits of data centers, along with potential
satellite and energy usage monitoring to flag unauthorized training
clusters.

\textbf{Key Takeaways for Building a Compute Pause Button and
Limitations}

\begin{itemize}
\item
  Dual-use tech governance can work when states have clear obligations,
  a technical implementation body, and verifiable limits.
\item
  Supply chain control over enrichment and reactor technology mirrors
  proposed mechanisms like traceable chip registries, chain-of-custody,
  and restricted export lists for high-performance compute.
\item
  Verification doesn't rely on trust alone: just like inspections and
  material accountancy in nuclear governance, our proposal includes
  auditable compute receipts, licensing logs, and location-verifiable
  chip deployments.
\item
  The model shows how limited access to dangerous capabilities can be
  traded for safe access to beneficial ones. This is a bargain that
  could inspire AI policy, which needs to balance safety concerns with
  the promotion of technological innovation and economic growth.
\end{itemize}

\textbf{Limitation:} The NPT-IAEA has struggled with enforcement against
powerful actors (such as North Korea, as discussed above) and thus has
not prevented all proliferation. However, most other nuclear-weapon
states have complied or been pressured to comply, and thus NPT-IAEA
appears to still be the most successful model we have for preventing
catastrophic nuclear technology misuse.

\hypertarget{policy-analogue-2-the-chemical-weapons-convention-cwc-and-the-organisation-for-the-prohibition-of-chemical-weapons-opcw}{%
\subsection{Policy Analogue 2: The Chemical Weapons Convention
(CWC) and the Organisation for the Prohibition of Chemical Weapons
(OPCW)}\label{policy-analogue-2-the-chemical-weapons-convention-cwc-and-the-organisation-for-the-prohibition-of-chemical-weapons-opcw}}

\textbf{Analogue Snapshot: Governing hard-to-detect threats through
supply control, on-site inspections, and multilateral enforcement}

The Chemical
Weapons Convention (CWC), enforced by the Organisation for the
Prohibition of Chemical Weapons (OPCW), is one of the strongest
examples of a global regime aimed at eliminating an entire class of
weapons considered too dangerous for any legitimate use. While chemical
weapons themselves are not dual-use, many of the precursor chemicals,
manufacturing equipment, and production facilities associated with them
do have legitimate civilian applications. This creates enforcement
challenges similar to those posed by advanced AI systems: the same
infrastructure used for beneficial purposes can, in the wrong hands, be
redirected toward catastrophic ends.

The CWC establishes the legal and political commitments to prohibit
chemical weapons, while the OPCW is the implementing body responsible
for monitoring compliance, conducting inspections, and coordinating
investigations. Together, they provide a model for how dual-use
precursor chemical materials and production facilities can be governed
through a blend of transparency, verification, and coordinated
enforcement. This model works even in cases where the harmful use of the
technology is hard to distinguish from legitimate applications (which
has similarities to the AI development context), and where production
activities can be easily concealed.

\textbf{GEV Mapping}

\textbf{Governance}

The CWC is a binding multilateral treaty ratified by 193 countries,
making it one of the most universally adopted arms control agreements.
It prohibits
the development, production, acquisition, stockpiling, retention,
transfer, and use of chemical weapons \parencite{organisationfortheprohibitionofchemicalweaponsArticleGeneralObligations}. State Parties commit to annual
declarations of relevant facilities, precursor materials, and equipment,
and accept the OPCW's authority to inspect and verify compliance.
Governance is maintained through regular review conferences, a governing
Executive Council, and consensus-based decision-making processes,
ensuring political legitimacy and adaptability over time.

In a compute governance framework, similar commitments could be
operationalized through a global compute supply registry, chip
production quotas, and mandatory declarations of data center
infrastructure, administered by a dedicated AI oversight body analogous
to the OPCW.

\textbf{Enforcement}

The CWC includes built-in provisions for enforcement through both collective
and individual action \parencite{chemicalweaponsconventionarchiveMainPositionsIssues}. If a violation is suspected or confirmed, the
OPCW can recommend measures to the Conference of States Parties or refer
the case to the United Nations Security Council. In practice,
enforcement has ranged from diplomatic pressure to international
sanctions \parencite{u.s.departmentofstateImposingNewMeasures}. While challenges exist in politically sensitive cases, the
treaty's normative force has led to meaningful disarmament: virtually
all declared chemical weapons stockpiles have been verifiably
destroyed \parencite{organisationfortheprohibitionofchemicalweaponsOPCWConfirmsAll}. The OPCW has also investigated high-profile uses of
chemical weapons in Syria and elsewhere, sometimes under difficult
political conditions.

Our Pause Button framework builds enforcement into the infrastructure
itself: offline compute licenses can be revoked, chips can be disabled
or throttled if used outside of approved parameters, and
self-deactivation mechanisms can be triggered in response to suspicious
activity or noncompliance. Additionally, governments could impose
sanctions for unauthorized chip acquisition or facility use, coordinated
through multilateral agreements.

\textbf{Verification}

The OPCW runs a rigorous verification regime, including: on-site
inspections (of declared production and storage facilities), monitoring
of dual-use chemicals via a ``Schedules''
system (which classifies substances based on potential for misuse) \parencite{organisationfortheprohibitionofchemicalweaponsAnnexChemicals},
requesting challenge
inspections (which allow member states to request surprise inspections
if noncompliance is suspected) \parencite{departmentofcommercebureauofindustryandsecurityChemicalWeaponsConvention}, sampling and analysis (where inspectors
can test materials for traces of chemical agents), and data
reconciliation and reporting (using annual declarations and audits of
trade and facility records).

This multilayered system makes it difficult for states to maintain
covert chemical weapons programs without detection, and provides clear
consequences for noncompliance. The system demonstrates how technical
monitoring and international inspection authority can be combined to
address a dispersed and often invisible, or hard-to-detect threat
landscape. This has key parallels to the risks posed by advanced AI
development. Thus we propose a similar multilayered verification system
for compute: chip-level telemetry, secure compute receipts,
cloud-provider usage logs linked to KYC identities, on-site audits of
major data centers, and the ability to request challenge inspections in
facilities suspected of exceeding compute thresholds.

\textbf{Key Takeaways for Building a Compute Pause Button and
Limitations}

\begin{itemize}
\item
  The CWC/OPCW regime shows how creating clear definitions, promoting
  shared norms, and coordinating verifiable commitments makes it
  possible to achieve near-global consensus on restricting a class of
  dangerous capabilities.
\item
  Its focus on precursor control offers a strong analogy for compute
  governance, where regulating access to compute (not just model
  outputs) may be the most practical way to limit capability escalation.
  This underlines the importance of tracking critical manufacturing
  tools, registering chip shipments, and limiting chip interconnect
  bandwidth to prevent scaling beyond agreed thresholds.
\item
  The OPCW's ability to conduct on-site inspections, verify facility
  use, and audit supply chains provides an important precedent for
  hardware-level verification and enforcement in AI. Our proposed
  toolkit includes challenge inspections, traceable chip registries, and
  facility compliance certifications to mirror these capabilities.
\end{itemize}

\textbf{Limitation:} The regime still faces enforcement challenges in
cases involving powerful
or non-cooperative actors \parencite{bureauofarmscontroldeterrenceandstabilityCondition10CAnnual2024}, especially when violations occur within
sovereign territory or in ongoing conflict zones. However, despite these
challenges, the OPCW remains a functioning and widely respected
institution. It demonstrates that even complex, technical, and
politically sensitive verification tasks can be carried out under
multilateral governance.

\hypertarget{coordinated-action-soft-law-norms-and-strategic-alignment}{%
\subsection{Coordinated Action: Soft Law, Norms, and Strategic
Alignment}\label{coordinated-action-soft-law-norms-and-strategic-alignment}}

The next two policy cases --- Operation Warp Speed and the Wassenaar
Arrangement --- offer insights into how coordination can emerge without
formal treaties, through soft law mechanisms, a strong sense of mission,
organizational agility, and the strategic use of incentives. These
models are relevant for shaping behavior in fast-moving,
economically-motivated, or politically fragmented domains, where trust
and urgency play an outsized role.

In the context of advanced AI, soft law models like these may serve as
critical transitional tools: helping coordinate early action, build
shared norms, and create momentum toward stronger treaty-based regimes.
They also offer agility in fast-moving domains, where formal governance
lags behind technological change. By aligning incentives, building
trust, and fostering institutional memory, soft law can lay the
foundation for more binding global arrangements in the future.

\hypertarget{policy-analogue-3-operation-warp-speed-ows}{%
\subsection{Policy Analogue 3: Operation Warp Speed
(OWS)}\label{policy-analogue-3-operation-warp-speed-ows}}

\textbf{Analogue Snapshot: Governing at the speed of innovation:
Accelerating deployment through agile public-private coordination}

Operation Warp Speed (OWS) was a U.S. government initiative launched in
2020 to accelerate the development, testing, production, and
distribution of COVID-19 vaccines. Within just one year, it helped
deliver multiple safe and effective vaccines, compressing a typical
5-10-year process into mere months. While its goals were very different
from AI safety, OWS offers a powerful precedent for how governments can
mobilize technical capacity, coordinate industry, and navigate
regulatory pathways, at speed --- without compromising quality. OWS
shows that under high-risk, high-urgency conditions, nations can build
fast, act decisively, and deliver safely, a lesson especially relevant
for AI, where rapid capability growth is already outpacing slow
institutional response.

\textbf{GEV Mapping}

\textbf{Governance}

Operation Warp Speed was governed by a cross-agency task force that
included the Department of Health and Human Services (HHS) (including
the CDC, FDA, NIH, and BARDA), the Department of Defense (DoD), and
private sector partners like Merck. It relied upon a centralized
command structure at the federal level capable of making high-level
strategic decisions across R\&D, logistics, procurement, distribution,
and regulation \parencite{slaouiDevelopingSafeEffective2020}. Rather than creating new legislation or agencies, OWS
used existing institutions with repurposed mandates, demonstrating that
\textbf{rapid coordination doesn't always require new bureaucracy}. The
governance of OWS was based on a clear mission and was given broad
discretion. Because the world faced extraordinary time pressure due to
the expanding pandemic, OWS was set up in a way that allowed it to cut
through red tape while maintaining accountability.

\textbf{Enforcement}

OWS is unique in that it did not rely on coercive enforcement but used
contractual incentives and conditional funding to shape industry
behavior. Pharmaceutical
companies received billions in upfront investment to develop vaccines,
but those funds were tied to clear milestones, regulatory requirements,
and performance targets \parencite{congressionalresearchserviceOperationWarpSpeed2021}. In effect, compliance was enforced through
financial leverage and federal purchasing power, rather than legal
compulsion. This `soft' enforcement model shows how governments can
guide and constrain private actors by structuring incentives wisely,
rather than simply outlawing unwanted behavior. This approach is
particularly relevant for compute governance, where frontier AI
development is driven by private companies; efforts to shape their
behavior may depend more on strategic incentives than on top-down
regulation. While our own compute governance framework emphasizes hard
legal mechanisms, the OWS example underscores that strategic use of
incentives and procurement power can play a vital complementary role.
Especially in early phases or politically sensitive contexts, soft
enforcement can help build compliance pathways and industry alignment
before stricter measures take hold.

\textbf{Verification}

Despite the speed, OWS preserved rigorous scientific and regulatory
oversight. The
FDA maintained its usual standards, and companies still had to pass all
three phases of clinical trials \parencite{u.s.food&drugadministrationEmergencyUseAuthorization2020}. What changed was parallelization:
manufacturing began before final approvals, and regulatory bodies were
involved throughout the R\&D process \parencite{slaouiDevelopingSafeEffective2020}. This created a system of
continuous, embedded verification, which ensured quality without delay. Independent
data monitoring boards, real-time trial data sharing, and early-stage
collaboration between regulators and firms provided transparency and
accountability \parencite{nationalinstituteofhealthNIHLaunchesClinical2020}. The process demonstrated that speed and safety need not
be in tension if verification is built into the process from the start.

\textbf{Key Takeaways for Building a Compute Pause Button and
Limitations}

\begin{itemize}
\item
  OWS demonstrates that governments can act swiftly in crisis:
  crucially, this speed was enabled by a shared sense of purpose,
  reinforced through federal coordination and strategic communication.
  In the context of AI, governance may serve not only to regulate but
  also to cultivate urgency, align incentives, and articulate a
  collective mission, all of which are essential for a timely and
  coordinated response to emerging risks.
\item
  Its model of centralized coordination, pre-emptive funding, and
  parallelized development and oversight offers some inspiration for
  fast-moving AI development and governance regimes. Remembering what
  was possible in the case of OWS can be especially valuable in
  crisis-response or high-alert scenarios involving potential
  catastrophic risk from dangerous AI.
\item
  Soft enforcement through strategic incentives (such as access to
  funding, regulatory cooperation, or critical infrastructure) could be
  used to guide behavior at the data center, chip, or model release
  levels, especially among frontier labs that depend on government
  goodwill, energy infrastructure, or market legitimacy.
\item
  Verification does not always require delay; real-time monitoring,
  embedded audits, and collaborative oversight could enable ongoing
  compliance checks for compute use.
\end{itemize}

Certain mechanisms of compute governance that we have proposed (such as
a global compute supply registry for advanced chips) could be launched
through voluntary coordination and soft-law incentives, creating early
visibility into compute flows without requiring immediate legal
mandates. Other transparency mechanisms, like KYC protocols for
large-scale cloud compute or voluntary audit disclosures, might also be
piloted in a soft-law context, helping establish norms and
infrastructure ahead of formal regulation.

\textbf{Limitation:} OWS was a national initiative under emergency
conditions with relatively aligned incentives (almost everyone wanted an
effective vaccine). Translating this model to global AI governance,
where actors have asymmetric interests and less shared urgency, will
require more sophisticated coordination and trust-building mechanisms.
Establishing a collective sense of mission in this context may depend on
strategic framing of risks, diplomatic agenda-setting, visible
leadership from coalitions of states, and the creation of focal points
(such as productive international summits or safety accords) that can
consolidate and deepen alignment around high-level goals.

\hypertarget{policy-analogue-4-the-wassenaar-arrangement}{%
\subsection{Policy Analogue 4: The Wassenaar
Arrangement}\label{policy-analogue-4-the-wassenaar-arrangement}}

\textbf{Analogue Snapshot: Coordinating export controls on dual-use
technologies through soft law and multilateral norm-setting}

The Wassenaar Arrangement on
Export Controls for Conventional Arms and Dual-Use Goods and
Technologies is a multilateral framework designed to prevent the
destabilizing accumulation of sensitive technologies by coordinating
national export controls. While it is not a binding treaty, Wassenaar
operates as a soft law regime by relying on transparency, peer pressure,
and shared norms rather than legal enforcement. It provides an
instructive analogue for AI compute governance - not as a complete
solution - but as a potential foundation for early-stage alignment,
especially where binding agreements are not yet politically feasible.
However, given the extreme risks and incentives around frontier AI, any
Wassenaar-like coordination would need to evolve into a much stronger
regime to be truly effective.

\textbf{GEV Mapping}

\textbf{Governance}

The Wassenaar Arrangement was established in 1996 and currently includes
42 participating states, including major exporters like the United
States, EU member states, Japan, and South Korea. It functions through
annual plenary meetings, technical expert groups, and working groups
that maintain and update control lists of dual-use goods and
technologies. Participation is voluntary, and decisions are made by
consensus. Each member implements the control lists through its own
national export laws, making governance decentralized but harmonized.
While not legally binding, Wassenaar's control lists have been widely
adopted into national regulation, including the \href{https://www.bis.gov/regulations/ear}{U.S. Commerce Control
List} and the \href{https://eur-lex.europa.eu/EN/legal-content/summary/dual-use-export-controls.html}{EU Dual-Use Regulation}.

\textbf{Enforcement}

As mentioned, there is no centralized enforcement body. Instead, each
participating state is responsible for enforcing its own export
controls. The arrangement depends on mutual
information exchange among member states and transparency where
participants share information on sensitive transfers and denials of
export licenses \parencite{armscontrolassociationWassenaarArrangementGlance2022}. This creates a network of informal accountability,
where reputational consequences and policy convergence drive compliance.
The soft enforcement approach is less coercive than treaty-based
regimes, but can still have real-world impact, where companies face
delayed approvals, increased scrutiny, or reputational risk for
violating norms, and governments may face diplomatic pressure for lax
controls.

\textbf{Verification}

There is no formal verification mechanism in Wassenaar; no inspectors,
audits, or surprise checks. Instead, information exchange and peer
coordination serve as proxies for trust. Members submit regular
notifications on sensitive technology transfers and export denials,
enabling pattern recognition and policy alignment.

\textbf{Key Takeaways for Building a Compute Pause Button and
Limitations}

\begin{itemize}
\item
  Wassenaar shows that even without a binding treaty or centralized
  authority, there can be an impact on supply chain governance through
  multilateral coordination, transparency, and soft enforcement.
\item
  The model may be especially relevant for compute, where export
  controls on chips, lithography equipment, or specialized data center
  hardware can act as effective pressure points for slowing or
  redirecting capability development.
\item
  Wassenaar's use of shared control lists which are periodically updated
  offers a blueprint for defining and updating thresholds for sensitive
  compute hardware.
\end{itemize}

\textbf{Limitation:} Wassenaar depends on consensus and national
discretion, which means enforcement and implementation can be uneven,
and major non-participants (e.g., China) are outside the regime. Its
soft-law nature limits its ability to handle actors actively seeking to
evade controls or operate outside the established supply chains. For
instance, some scholars have argued that the Arrangement has not had a noticeable effect on
weapon sales because of strong economic incentives to continue such
sales \parencite{lewis2015effectiveness}.

\textbf{Policy Analogues Conclusion}

These policy analogues demonstrate that global governance of powerful,
dual-use technologies is possible and supported by real-world
precedents, even when risks are high and state interests diverge. The
binding regimes (NPT/IAEA and CWC/OPCW) show the enduring value of hard
law: clear commitments, formal oversight, and enforceable verification.
The softer coordination models (Operation Warp Speed and the Wassenaar
Arrangement) reveal how strategic incentives, urgency, and norm-building
can also drive meaningful alignment, especially under time pressure or
political constraints.

For AI compute governance, these lessons suggest that a credible Compute
Pause Button will require a loosely hybrid, layered institutional
approach: one that combines the rigor of legal frameworks with the
flexibility and early adoption of soft power tools. Even where binding
treaties are difficult or may take time to institute, progress can be
made by strengthening shared incentives, building institutional trust,
and establishing coordination mechanisms to keep pace with rising risks
and capabilities.

\hypertarget{conclusion-engineering-the-pause-button-together}{%
\section{Conclusion: Engineering the Pause Button
Together}\label{conclusion-engineering-the-pause-button-together}}

This report proposes a concrete and enforceable architecture for a
global Compute Pause Button: one capable of preventing frontier AI
training runs that exceed agreed safety thresholds. At the heart of this
architecture is a framework that aligns three essential intervention
points --- technical containment, traceability infrastructure, and
regulatory authority --- with a structured system of Governance,
Enforcement, and Verification (GEV).

By organizing existing and proposed mechanisms within this integrated
framework, we offer a practical roadmap for making a pause not just
imaginable, but implementable. Unlike proposals that focus solely on
model evaluations or post-hoc oversight, this approach targets the
material infrastructure (e.g., chips, clusters, and supply chains) that
enables dangerous scale. It builds on precedents from nuclear
non-proliferation, chemical weapons control, financial regulation, and
emergency mobilization, but adapts them to the unique speed and
complexity of AI development.

We recognize that realizing this vision will depend on critical
assumptions: that political will can be mobilized, that technical
safeguards can mature in time, that international cooperation is
achievable, and that civil society will demand action. This report does
not resolve those dependencies, but it is designed to support them by
clarifying what a pause would require, and where credible points of
control already exist.

The path ahead will require institutional imagination, coordinated
experimentation, and unprecedented cooperation across governments,
industry, and the research community. But the alternative --- a world
where the scale of AI development escapes our ability to steer it ---
demands no less.

We call on policymakers, AI labs, chipmakers, and multilateral
institutions to begin coordinated planning around compute governance
today. The architecture for a global pause is within reach. What is
needed now is the collective will to build it.

\section*{Acknowledgments}
This project was conducted under the auspices of the 10th AI Safety Camp (AISC), which convened and supported our research team. We thank AISC for sponsoring this work and for fostering a collaborative environment for advancing critical questions in AI governance.

\printbibliography

\end{document}